\documentclass[aps,prd,amsmath]{revtex4}
\usepackage{CJK}
\usepackage{amsmath}
\usepackage{enumerate}
\usepackage{amssymb}
\usepackage{amsfonts}
\usepackage{mathrsfs}
\usepackage{latexsym}
\usepackage{bm}
\usepackage{amscd}
\usepackage{graphicx}
\usepackage{epsfig}
\usepackage{hhline,multirow}
\usepackage{dcolumn}
\usepackage{url}
\usepackage[colorlinks=true,linkcolor=blue]{hyperref}
\usepackage{color}
\usepackage{indentfirst}
\usepackage{subfigure}
\usepackage{psfrag}
\usepackage{slashed}
\usepackage{float}
\usepackage{setspace}
\begin{document}
\title{Contributions to the nucleon form factors from bubble and tadpole diagrams}
\author{Z. Y. Gao}
\affiliation{Institute of High Energy Physics, CAS, P. O. Box 918(4), Beijing 100049, China}
\affiliation{School of Physical Sciences, University of Chinese Academy of Sciences, Beijing 100049, China}
\author{P. Wang}
\affiliation{Institute of High Energy Physics, CAS, P. O. Box 918(4), Beijing 100049,
China}
\affiliation{School of Physical Sciences, University of Chinese Academy of Sciences, Beijing 100049, China}
\author{M. Y. Yang}
\affiliation{Institute of High Energy Physics, CAS, P. O. Box 918(4), Beijing 100049,
China}
\begin{abstract}
The nonlocal chiral effective theory is applied to investigate the electromagnetic and strange form factors of nucleon. The bubble and tadpole diagrams are included in the calculation. With the contributions from bubble and tadpole diagrams, the obtained electromagnetic form factors are close to the results without these contributions as long as the low energy constants $c_1$ and $c_2$ are properly chosen, while the magnitudes of strange form factors become larger. The electromagnetic form factors are  in good agreement with the experimental results, while the magnitudes of strange form factors are larger than the lattice data.
\end{abstract}
\maketitle

\section{Introduction}

The electromagnetic form factors of nucleon are important physical quantities which are helpful to understand both the nucleon inner structure and the mechanism of strong interaction. For many years the elastic lepton-nucleon scattering measurements have enabled us to get detailed information on the electromagnetic form factors of the proton and neutron over a wide range of kinematics. A lot of measurements were carried out at the Stanford Mark III accelerator \cite{Janssens}, the Cambridge Electron Accelerator \cite{Price}, the Stanford Linear Accelerator Center (SLAC) \cite{Lung}, Bonn \cite{Bruins}, DESY \cite{Bartel}, Mainz \cite{Bernauer}, NIKHEF \cite{Anklin}, MIT-Bates \cite{Markowitz}, and at Jefferson Lab \cite{Lachniet}. The elastic scattering measurements have stimulated considerable activity over the past two decades in the determination of the flavor separated form factors of the up, down and strange quarks in the nucleon. A number of measurements for the strange form factors have been successfully performed, starting with SAMPLE at Bates \cite{Spayde} and A4 at Mainz \cite{Maas}, followed by the high precision G0 \cite{Armstrong} and HAPPEX \cite{Acha} experiments at Jefferson Lab. 

There are also numerous theoretical work on the nucleon form factors. Due to the non-perturbative property of the strong interaction, it is very difficult to study hadron properties using the fundamental theory QCD directly.
Many theoretical calculations are based on the phenomenological methods, such as the constituent quark model, cloudy bag model, perturbative chiral quark model, quark-diquark model, nonlocal quark meson coupling model, Nambu-Jona-Lasinio model, Dyson-Schwinger equations, vector meson dominance model, chiral quark soliton model, and AdS/QCD approach, etc \cite{Gross,Zhang,Dai,Thomas,Lu,Lyubovitskij,Sanctis,Ivanov,Faessler,Weigel,Serrano,Roberts,Bijker,Brodsky,Chen,Schweitzer,Goeke}.
Lattice QCD simulation is the most rigorous method to study hadron properties. However, the simulated results have to be extrapolated to the continuum and infinite volume limits in order to compare with experimental data. In addition, if the simulations are performed using heavy quark masses, the data need to be extrapolated to the physical quark masses.
With the improvement of the computing speed, currently, lattice simulation on the nucleon form factors can be carried out at physical quark masses including the disconnected contributions \cite{Sufian,Jang}.

Besides lattice QCD, chiral effective field theory (EFT) is another systematic approach which has been widely applied in hadron physics \cite{Weinberg,Gasser}. Compared with the various phenomenological models, one advantage of effective theory is that it can provide some model independent results. Although Chiral EFT has been a fairly successful approach, for the nucleon form factors it is only valid at relatively small $Q^2$ values, say less than 0.1 GeV$^2$ \cite{Fuchs}. The range can be extended up to 0.4 GeV$^2$ by explicitly including vector meson degrees of freedom into the theory \cite{Kubis}.

In recent years, we proposed a nonlocal chiral effective theory, which makes it possible to study hadron properties at relatively large momentum transfer \cite{Wang1}. The basic idea is that in the nonlocal Lagrangian, baryon and meson are located in different coordinates described by a correlator. To guarantee the local gauge invariance, the path integral of the gauge field is introduced which leads to the additional interactions. The additional Feynman diagrams are crucial to get the corrected normalized nucleon charge. On the one hand, the correlator makes the loop integral ultraviolet convergent. On the other hand, the physical quantities at large momentum transfer can be described very well. Perturbative calculations in chiral EFT expand observables as series in the pseudoscalar meson mass ${\cal O}(m_\phi/\Lambda_{\chi})$ or small external momentum ${\cal O}(q/\Lambda_{\chi})$, where $\Lambda_{\chi} \sim 1$ GeV is the scale associated with the chiral EFT. With the introducing of the regulator, the power counting of our nonlocal effective field theory is lost. The higher order terms are included as a kind of resummation. This resummation of the chiral expansion induced through the introduction of a finite range cutoff in the momentum integrals of meson-loop diagrams has been shown to be a very good and effective resummation method.
The nonlocal chiral EFT has been applied to study the nucleon electromagnetic form factors \cite{He1}, strange form factors \cite{He2}, unpolarized and polarized parton distribution functions (PDFs) \cite{Salamu1,Salamu2,He3} and generalized parton distributions (GPDs) \cite{He4}, etc. 
The approach has also been generalized to the fundamental interaction of QED, which provided an interesting explanation to the lepton $g-2$ anomaly \cite{He5,Li}. 

In the previous form factor calculations, the bubble and tadpole diagrams were not included \cite{He1,He2,Yang}. In fact, in the earlier extrapolation of nucleon form factors with finite-range regularization, these diagrams were not include either \cite{Wang2,Wang3}. The bubble and tadpole diagrams have no effect on the nucleon wave function renormalization because their contributions to the nucleon charge are summed to be zero. In this paper, we will apply the nonlocal EFT to study the nucleon form factors including these bubble and tadpole diagrams to see whether the results are changed or not.
The paper is organized in the following way. In section II, we briefly introduce the nonlocal chiral effective Lagrangian. 
The matrix elements of the bubble and tadpole diagrams for the electromagnetic and strange form factors are presented in section III. Numerical results are discussed in section IV. Finally, section V is a brief summary. 

\section{Chiral Effective Lagrangian}

We start from the local chiral effective Lagrangian. The lowest Lagrangian which describes the pseudoscalar mesons, octet and decuplet baryons and their interaction is expressed as 
\begin{eqnarray}
\mathcal{L}^{(1)} & = & {\rm Tr}\big[\bar{B}\left(i\slashed{\mathcal{D}}-M_{B}\right)B\big]+D\,{\rm Tr}\left[\bar{B}\gamma^{\mu}\gamma^{5}\left\{ u_{\mu},B\right\} \right]+F\,{\rm Tr}\left[\bar{B}\gamma^{\mu}\gamma_{5}\left[u_{\mu},B\right]\right]\nonumber \\
 &  & +\overline{T}_{\mu}^{ijk}\left(i\gamma^{\mu\nu\alpha}\mathcal{D}_{\alpha}+iM_{T}\sigma^{\mu\nu}\right)T_{\nu}^{ijk}+\mathcal{C}\left[\epsilon^{ijk}\overline{T}_{\mu}^{ilm}\Theta^{\mu\nu}\left(u_{\nu}\right)^{lj}B^{mk}+\text{h.c.}\right] \nonumber \\ 
 &  & +\mathcal{H}\,\overline{T}_{\mu}^{ijk}\gamma^{\mu\nu\alpha}\gamma^{5}\left(u_{\alpha}\right)^{kl}T_{\nu}^{ijl}+\frac{f^{2}}{4}{\rm Tr}\big[\partial_{\mu}U\left(\partial^{\mu}U\right)^{\dagger}\big],
\end{eqnarray}
where $D$, $F$, ${\cal C}$ and $\mathcal{H}$ are the baryon-meson coupling constants. $M_{B}$ and $M_{T}$ are the octet and decuplet
baryon masses. $f\thickapprox93$ MeV is the pseudoscalar decay
constant. The $\Theta^{\mu\nu}$ in octet-decuplet transition operator is given by 
\begin{equation}
\Theta^{\mu\nu}=g^{\mu\nu}-Z\gamma^{\mu}\gamma^{\nu},
\end{equation}
where $Z$ is the off shell parameter. 
The tensors $\gamma^{\mu\nu\alpha}$ is defined as  $\gamma^{\mu\nu\alpha}=-i\left\{ \sigma^{\mu\nu},\gamma^{\alpha}\right\}$.
The octet baryons and mesons are arranged in the $3\times 3$ matrices, while the decuplet baryons are represented by the symmetric tensor with three indices.
The covariant derivatives of the octet and decuplet baryon
fields are given by
\begin{gather}
\mathcal{D}_{\mu}B=\partial_{\mu}B+[\Gamma_{\mu},B], \nonumber\\
\mathcal{D_{\mu}}T_{\nu}^{ijk}=\partial_{\mu}T_{\nu}^{ijk}+(\Gamma_{\mu},T_{\nu})^{ijk}. 
\end{gather}
The mesons couple to the baryon fields through the vector and axial vector combinations defined as 
\begin{align}
 & \Gamma_{\mu}=\frac{1}{2}(u\partial_{\mu}u^{\dagger}+u^{\dagger}\partial_{\mu}u), \nonumber \\
 & u_{\mu}=\frac{i}{2}(u\partial_{\mu}u^{\dagger}-u^{\dagger}\partial_{\mu}u),
\end{align}
where ${u}$ is defined in terms of the pseudoscalar meson field
${\phi}$
\begin{equation}
u^{2}=\exp(i\frac{\sqrt{2}\phi}{f}).
\end{equation}
The octet, decuplet and octet-decuplet transition magnetic
moment operators are needed in the one loop calculation of nucleon form factors. The magnetic Lagrangian is written as 
\begin{align}\label{eq:mag}
{\cal L}_{\text{mag}} & =\frac{e}{4M_{B}}(c_{1}\textrm{Tr}[\overline{B}\sigma^{\mu\nu}\{\mathcal{F}_{\mu\nu}^{\dagger},B\}]+c_{2}\textrm{Tr}[\overline{B}\sigma^{\mu\nu}[\mathcal{F}_{\mu\nu}^{\dagger},B]]+c_{3}\textrm{Tr}[\mathcal{F}_{\mu\nu}^{\dagger}]\textrm{Tr}[\overline{B}\sigma^{\mu\nu}B])\nonumber \\
 & -\frac{eF_{2}^{T}}{4M_{T}}\overline{T}_{\mu}^{abc}\sigma^{\alpha\beta}\mathcal{F}_{\alpha\beta}Q^{ce}T^{\mu,eba} \nonumber\\
 & +\frac{iec_{4}}{4M_{B}}\mathcal{F}_{\mu\nu}(\epsilon^{ijk}\overline{B}^{jm}Q^{il}\gamma^{\mu}\gamma^{5}T^{\nu,klm}+\epsilon^{ijk}\overline{T}^{\mu,klm}Q^{li}\gamma^{\nu}\gamma^{5}B^{mj}), \end{align}
where $c_{1}$, $c_{2}$ and $c_{3}$ describe the magnetic moments of octet baryons at tree level, while $F_2^T$ ad $c_4$ are related to the magnetic moments of decuplet baryons and octet-decuplet transition. With the SU(6) symmetry, one can have
the following relationships \cite{He4}
\begin{align}
c_3 & = c_2 - c_1, \nonumber \\
F_{2}^T & =\frac{1}{3}(c_{1}+3c_{2}), \nonumber \\
c_4 & = \frac{1}{\sqrt{3}}c_{1}.
\end{align}
To calculate the contributions of the bubble diagrams, the following high order Lagrangian which provides the next-to-leading order baryon-baryon-meson-meson interaction has to be included \cite{Kubis2}
\begin{equation}\label{eq:hiin}
\mathcal{L}^{(2)}=\frac{i}{2}\sigma^{\mu\nu}(b_{9}\textrm{Tr}[\overline{B}u_{\mu}]\textrm{Tr}[u_{\nu}B]+b_{10}\textrm{Tr}[\overline{B}\{[u_{\mu},u_{\nu}],B\}]+b_{11}\textrm{Tr}[\overline{B}[[u_{\mu},u_{\nu}],B]]),
\end{equation}
where the coefficients $b_{9}$, $b_{10}$ and $b_{11}$ are determined in Ref.~\cite{Kubis2} as $b_{9}=1.36$ GeV, $b_{10}=1.24$ GeV, and $b_{11}=0.46$ GeV. 

The gauge invariant nonlocal Lagrangian can be obtained by the displacement of the coordinates of the meson and photon fields with the proper inclusion of the gauge link. For example,
the gauge invariant local strong interaction $pp\pi\pi$ is expressed as
\begin{align}
{\cal L}_{\pi\pi}^{\text{local}}=\frac{i}{4f^{2}}\overline{p}(x)\gamma^{\mu}p(x)\left[\pi^{+}(x)(\partial_\mu -ieA_\mu(x))\pi^{-}(x)-\pi^{-}(x)(\partial_\mu+ieA_\mu(x))\pi^{+}(x)\right].
\end{align}
The corresponding nonlocal Lagrangian is expressed as 
\begin{eqnarray}\label{eq:nlpipi}
{\cal L}_{\pi\pi}^{\text{nl}} & = & \frac{i}{4f^{2}}\overline{p}(x)\gamma^{\mu}p(x)\int da\int dbF(a)F(b) \Big[\text{exp}\Big[ie\int_{x}^{x+a}dz_{\nu}\int dc\mathcal{A}^{\nu}(z-c)F(c)\Big]\pi^{+}(x+a)\nonumber \\
 & \times & (\partial_{\mu}-ie\int dcF(c)\mathcal{A}_{\mu}(x-c))\text{exp}\Big[-ie\int_{x}^{x+b}dz_{\nu}\int dc\mathcal{A}^{\nu}(z-c)F(c)\Big]\pi^{-}(x+b)\nonumber\\
 & - & (\partial_{\mu}+ie\int dcF(c)\mathcal{A}_{\mu}(x-c))\text{exp}\Big[ie\int_{x}^{x+a}dz_{\nu}\int dc\mathcal{A}^{\nu}(z-c)F(c)\Big]\pi^{+}(x+a)\nonumber \\
 & \times & \text{exp}\Big[-ie\int_{x}^{x+b}dz_{\nu}\int dc\mathcal{A}^{\nu}(z-c)F(c)\Big]\pi^{-}(x+b)\Big],
\end{eqnarray}
where $F(a)$ is the correlation function. To guarantee the gauge
invariance, the gauge links are introduced in the above Lagrangian. 
Similarly, the nonlocal electromagnetic interaction of nucleon can also be obtained. For example, the local interaction between proton and photon is written as 
\begin{align}
{\cal L}_{\text{EM}}^{\text{local}}= & -e\bar{p}(x)\gamma^{\mu}p(x)\mathcal{A}_{\mu}(x)+\frac{(c_{1}+3c_{2})e}{12m_{N}}\bar{p}(x)\sigma^{\mu\nu}p(x)F_{\mu\nu}(x).
\end{align}
The corresponding nonlocal Lagrangian is expressed as
\begin{align}
{\cal L}_{\text{EM}}^{\text{nl}}=-e\int da\bar{p}(x)\gamma^{\mu}p(x)\mathcal{A}_{\mu}(x-a)F_{1}(a)+\frac{(c_{1}+3c_{2})e}{12m_{N}}\int da\bar{p}(x)\sigma^{\mu\nu}p(x)F_{\mu\nu}(x-a)F_{2}(a),
\end{align}
where $F_{1}(a)$ and $F_{2}(a)$ are the correlation functions for the nonlocal electric and magnetic interactions. As in our previous work, we assume that the charge and magnetic form factors at tree level have the same momentum dependence as nucleon-pion vertex, i.e.
$G_{M}^{{\rm tree}}(p)=\mu_p G_{E}^{{\rm tree}}(p)=\mu_p \tilde{F}(p)$,
where $\tilde{F}(p)$ is the Fourier transformation of the correlation function $F(a)$. Therefore, the corresponding functions of $\tilde{F}_{1}(q)$ and $\tilde{F}_{2}(q)$ are then expressed as 
\begin{eqnarray}
\tilde{F}_{1}(q) & = & \tilde{F}(q)\frac{4m_{N}^{2}+\mu_{p}Q^{2}}{4m_{N}^{2}+Q^{2}},~~~\tilde{F}_{2}(q)=\tilde{F}(q)\frac{(\mu_{p}-1)4m_{N}^{2}}{4m_{N}^{2}+Q^{2}}.
\end{eqnarray}
The nonlocal Lagrangian is invariant under the following gauge transformation 
\begin{eqnarray}
\pi^{+}(y)\rightarrow e^{i\alpha(y)}\pi^{+}(y),~~~~p(x)\rightarrow e^{i\alpha(x)}p(x),~~~~\mathcal{A}_{\mu}(x)\rightarrow\,\mathcal{A}_{\mu}(x)-\frac{1}{e}\partial_{\mu}\alpha'(x),
\end{eqnarray}
where $\alpha(x)=\int\!\,da\alpha^{\prime}(x-a)F(a)$.

From the nonlocal Lagrangian, one can get the interaction with the external photon. There are two kind of electromagnetic interactions in nonlocal case. Besides the one from the minimal substitution which is the same as in the local interaction, there is an additional one generated from the expansion of the gauge link. For example, from Eq.~(\ref{eq:nlpipi}), we can get the additional interaction as
\begin{align}
{\cal L}^{\text{add}} & =-\frac{e}{4f^{2}}\overline{p}(x)\gamma^{\mu}p(x)\int da\int dbF(a) F(b)\Big[\int_{x}^{x+a}dz_{\nu}\int dc\mathcal{A}^{\nu}(z-c)F(c)\pi^{+}(x+a)\partial_{\mu}\pi^{-}(x+b)\nonumber \\
 & -\pi^{+}(x+a)\partial_{\mu}\Big(\int_{x}^{x+b}dz_{\nu}\int dc\mathcal{A}^{\nu}(z-c)F(c)\pi^{-}(x+b)\Big)\nonumber \\
 & -\partial_{\mu}\Big(\int_{x}^{x+a}dz_{\nu}\int dc\mathcal{A}^{\nu}(z-c)F(c)\pi^{+}(x+a)\Big)\pi^{-}(x+b)\nonumber \\
 & +\partial_{\mu}\pi^{+}(x+a)\int_{x}^{x+b}dz_{\nu}\int dc\mathcal{A}^{\nu}(z-c)F(c)\pi^{-}(x+b)\Big].
\end{align}
The additional interaction is important to get the renormalized proton (neutron) charge 1 (0). With the nonlocal Lagrangian, one can calculate the electromagnetic and strange form factors of nucleon.

\section{Electromagnetic Form Factors}

\begin{figure}
\centering{}\includegraphics[scale=0.85]{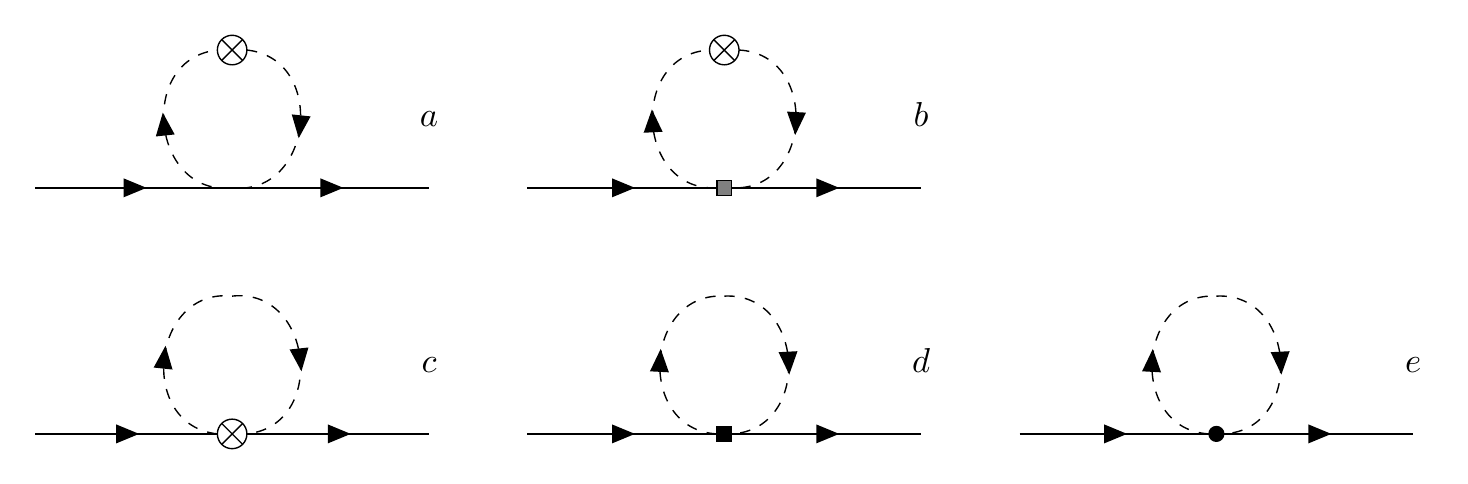} \caption{One-loop Feynman diagrams for the nucleon electromagnetic form factors. The solid and dashed lines are for nucleon and meson, respectively. The crossed circle represents the interaction with photon field from the minimal substitution, black square represents the magnetic interaction in Eq.~(\ref{eq:mag}), gray square denotes the interaction in Eq.~(\ref{eq:hiin}), and filled circle denotes additional gauge link interaction with the photon field.}
\label{diagrams} 
\end{figure}

The Dirac and Pauli form factors of nucleon are defined as 
\begin{equation}
<N(p')|J^{\mu}|N(p)>=\bar{u}(p')\left\{ \gamma^{\mu}F_{1}^{N}(Q^{2})+\frac{i\sigma^{\mu\nu}q_{\nu}}{2m_{N}}F_{2}^{N}(Q^{2})\right\} u(p),\label{eq:f1f2}
\end{equation}
where $q=p^{\prime}-p$ and $Q^{2}=-q^{2}$. $F_{1}^{N}(Q^{2})$ and $F_{2}^{N}(Q^{2})$ are the Dirac and Pauli form factors, respectively. The combination
of the above form factors can generate the electric and magnetic form factors as 
\begin{eqnarray}
G_{E}^{N}(Q^{2})=F_{1}^{N}(Q^{2})-\frac{Q^{2}}{{4m_{N}^{2}}}F_{2}^{N}(Q^{2})\,\,\,\,\,\,\,\,\,\,\,\,\,\,G_{M}^{N}(Q^{2})=F_{1}^{N}(Q^{2})+F_{2}^{N}(Q^{2}).
\end{eqnarray}
In Eq.~(\ref{eq:f1f2}), the contributions of the $u$, $d$ and $s$ quark are all included in the current. If only strange current is included, one can get the strange form factors. 
According to the Lagrangian, the one-loop Feynman diagrams which contribute to the nucleon electromagnetic form factors are plotted in Fig.~\ref{diagrams}, where only bubble and tadpole diagrams are shown. In Fig.~\ref{diagrams}, diagrams $a$ and $b$ are called bubble diagrams, while $c$, $d$ and $e$ are called tadpole diagrams. Here, we only present the formulas for the bubble and tadpole diagrams. The formulas for the other diagrams can be found in Refs.~\cite{He1,Yang}.

The contributions of the leading bubble diagram Fig.~\ref{diagrams}a to the matrix elements in Eq.~(\ref{eq:f1f2}) are written as
\begin{align}
\Gamma_{a}^{\mu(p)} & =-\frac{1}{4f^{2}}I^\mu_{a,\pi^{+}\pi^{-}}-\frac{1}{2f^{2}}I^\mu_{a,K^{+}K^{-}},\nonumber \\
\Gamma_{a}^{\mu(n)} & =\frac{1}{4f^{2}}I^\mu_{a,\pi^{+}\pi^{-}}-\frac{1}{4f^{2}}I^\mu_{a,K^{+}K^{-}},
\nonumber\\
\Gamma_{a}^{\mu(s)} & =\frac{1}{4f^{2}}I_{a,K^{+}K^{-}}^{\mu}+\frac{1}{2f^{2}}I_{a,K^{0}\bar{K}^0}^{\mu},
\end{align}
where the integral $I_{a,\phi^+\phi^-}^{\mu}$ is expressed as
\begin{equation}
I^\mu_{a,\phi^+\phi^-}=\tilde{F}(q)\,\bar{u}(p')\int\!\frac{d^{4}k}{(2\pi)^{4}}\tilde{F}(k)\frac{1}{D_\phi(k)}(2\slashed{k}+\slashed{q})2k^{\mu}\tilde{F}(k+q)\frac{1}{D_\phi(k+q)}u(p).
\end{equation}
Besides the matrix elements for proton and neutron, the matrix element of strange quark $\Gamma_{a}^{\mu(s)}$ is also written.
In the above integral, $D_\phi(k)$ is the propagator of meson $\phi$. The correlators $\tilde{F}(k)$ and $\tilde{F}(k+q)$ make the loop integral convergent and $\tilde{F}(q)$ gives proper meson form factor at tree level. 
The contributions of the bubble diagram at next-to-leading order Fig.~\ref{diagrams}b to the matrix elements
are written as
\begin{align}
\Gamma_{b}^{\mu(p)} & =-\frac{2(b_{10}+b_{11})}{f^{2}}I^\mu_{b,\pi^{+}\pi^{-}}-\frac{4b_{11}+b_{9}}{f^{2}}I^\mu_{b,K^{+}K^{-}}, \nonumber \\
\Gamma_{b}^{\mu(n)} & =\frac{2(b_{10}+b_{11})}{f^{2}}I^\mu_{b,\pi^{+}\pi^{-}}-\frac{2(b_{10}-b_{11})}{f^{2}}I^\mu_{b,K^{+}K^{-}},\nonumber\\
\Gamma_{b}^{\mu(s)} & =-\frac{2(b_{10}-b_{11})}{f^{2}}I_{b,K^{+}K^{-}}^{\mu}+\frac{4b_{11}+b_{9}}{f^{2}}I_{b,K^{0}\bar{K}^0}^{\mu},
\end{align}
where the integral $I_{b,\phi^+\phi^-}^{\mu}$ is expressed as
\begin{equation}
I^\mu_{b,\phi^{+}\phi^{-}}=\tilde{F}(q)\,\bar{u}(p')\int\!\frac{d^{4}k}{(2\pi)^{4}}\sigma^{\rho\nu}q_{\rho}k_{\nu}\tilde{F}(k)\frac{1}{D_\phi(k)}2k^{\mu}\tilde{F}(k+q)\frac{1}{D_\phi(k+q)}u(p).
\end{equation}
For the tadpole diagram with electric coupling to the photon field, Fig.~\ref{diagrams}c, the matrix elements are written as
\begin{align}
\Gamma_{c}^{\mu(p)} & =-\frac{4m_{N}^{2}+Q^{2}(1+c_{1}+c_{2})}{2(4m_{N}^{2}+Q^{2})f^{2}}I_{c,\pi^{+}\pi^{-}}^{\mu}-\frac{4m_{N}^{2}+Q^{2}(1+c_{2})}{(4m_{N}^{2}+Q^{2})f^{2}}I_{c,K^{+}K^{-}}^{\mu}, \nonumber\\
\Gamma_{c}^{\mu(n)} & =\frac{4m_{N}^{2}+Q^{2}(1+c_{1}+c_{2})}{2(4m_{N}^{2}+Q^{2})f^{2}}I_{c,\pi^{+}\pi^{-}}^{\mu}-\frac{4m_{N}^{2}+Q^{2}(1-c_{1}+c_{2})}{2(4m_{N}^{2}+Q^{2})f^{2}}I_{c,K^{+}K^{-}}^{\mu},\nonumber\\
\Gamma_{c}^{\mu(s)} & =\frac{4m_{N}^{2}+Q^{2}(1-c_{1}+c_{2})}{2(4m_{N}^{2}+Q^{2})f^{2}}I_{c,K^{+}K^-}^{\mu}+\frac{4m_{N}^{2}+Q^{2}(1+c_{2})}{(4m_{N}^{2}+Q^{2})f^{2}}I_{c,K^{0}\bar{K}^0}^{\mu},
\end{align}
where the integral $I_{c,\phi^+\phi^-}^{\mu}$ is expressed as
\begin{equation}
I_{c,\phi^+\phi^-}^{\mu}=\tilde{F}(q)\,\bar{u}(p')\int\!\frac{d^{4}k}{(2\pi)^{4}}\tilde{F}^{2}(k)\frac{1}{D_{\phi}(k)}\gamma^{\mu}u(p).
\end{equation}
For the tadpole diagram with the magnetic coupling to the photon field, Fig.~\ref{diagrams}d, the matrix elements are written as
\begin{align}
\Gamma_{d}^{\mu(p)} & =-\frac{2m_{N}^{2}(c_{1}+c_{2})}{(4m_{N}^{2}+Q^{2})f^{2}}I_{d,\pi^+\pi^-}^{\mu}-\frac{4m_{N}^{2}c_{2}}{(4m_{N}^{2}+Q^{2})f^{2}}I_{d,K^{+}K^-}^{\mu}, \nonumber\\
\Gamma_{d}^{\mu(n)} & =\frac{2m_{N}^{2}(c_{1}+c_{2})}{(4m_{N}^{2}+Q^{2})f^{2}}I_{d,\pi^{+}\pi^-}^{\mu}+\frac{2m_{N}^{2}(c_{1}-c_{2})}{(4m_{N}^{2}+Q^{2})f^{2}}I_{d,K^{+}K^-}^{\mu},\nonumber\\
\Gamma_{d}^{\mu(s)} & =-\frac{2m_{N}^{2}(c_{1}-c_{2})}{(4m_{N}^{2}+Q^{2})f^{2}}I_{d,K^{+}K^-}^{\mu}+\frac{4m_{N}^{2}c_{2}}{(4m_{N}^{2}+Q^{2})f^{2}}I_{d,K^{0}\bar{K}^0}^{\mu},
\end{align}
where the integral $I_{d,\phi^+\phi^-}^{\mu}$ is expressed as
\begin{equation}
I_{d,\phi^+\phi^-}^{\mu}=\tilde{F}(q)\,\bar{u}(p')\int\!\frac{d^{4}k}{(2\pi)^{4}}\tilde{F}^{2}(k)\frac{1}{D_{\phi}(k)}\frac{\sigma^{\mu\nu}q_{\nu}}{2m_{N}}u(p).
\end{equation}
In the nonlocal chiral EFT, there is an additional interaction with the external photon, which is generated from the expansion of the gauge link. The contributions of this additional diagram Fig.~\ref{diagrams}e to the matrix elements are written as
\begin{align}
\Gamma_{e}^{\mu(p)} & =-\frac{1}{2f^{2}}I_{e,\pi^+\pi^-}^{\mu}-\frac{1}{f^{2}}I_{e,K^{+}K^-}^{\mu}, \nonumber\\
\Gamma_{e}^{\mu(n)} & =\frac{1}{2f^{2}}I_{e,\pi^+\pi^-}^{\mu}-\frac{1}{2f^{2}}I_{e,K^{+}K^-}^{\mu},
\nonumber\\
\Gamma_{e}^{\mu(s)} & =\frac{1}{2f^{2}}I_{e,K^{+}K^-}^{\mu}+\frac{1}{f^{2}}I_{e,K^{0}\bar{K}^0}^{\mu},
\end{align}
where the integral $I_{e,\phi^+\phi^-}^{\mu}$ is expressed as
\begin{equation}
I_{e,\phi^+\phi^-}^{\mu}=\tilde{F}(q)\,\bar{u}(p')\int\!\frac{d^{4}k}{(2\pi)^{4}}\tilde{F}(k)\frac{1}{D_{\phi}(k)}2\slashed{k}\frac{(2k+q)^{\mu}}{2k\cdot q+q^{2}}[\tilde{F}(k+q)-\tilde{F}(k)]u(p).
\end{equation}
After simplifying the $\gamma$ matrix algebra, we can get the separate expressions for the Dirac and Pauli form factors. 
Together with the contributions from the rainbow and Kroll-Rudderman diagrams which have been obtained in our previous work \cite{He1,Yang}, the nucleon form factors can be obtained. In the next section, numerical results will be discussed.

\begin{table}[tbph]
\caption{The obtained parameters $c_{1}$ and $c_{2}$. The first and second two $c_i$ $(i=1,2)$ are for the cases without and with bubble and tadpole diagrams, respectively.}
\begin{tabular}{|c|c|c|c|c|}
~$\Lambda$ (GeV)~ & $c_{1}$ & $c_{2}$ & $c_{1}$ & $c_{2}$ \\
\hline 
0.9 & ~~2.05~~ & ~~0.75~~ & ~~1.77~~ & ~~0.67~~ \\
\hline 
1.0 & 1.98 & 0.74 & 1.61 & 0.63 \\
\hline 
1.1 & 1.92 & 0.72 & 1.46 & 0.59 \\
\end{tabular}
\end{table}

\section{Numerical Results}

In the numerical calculations, the parameters are chosen as $D=0.75$, $F=\frac{2}{3}D$, $\mathcal{H}=-3D$. The coupling constant ${\cal C}$ is chosen to be $1.0$ which
is as same as Refs.~\cite{He1,Pascalutsa}. 
The parameters $c_{1}$ and $c_{2}$ are determined by the experimental magnetic moments of proton and neutron $\mu_p=2.79$ and $\mu_n=-1.91$. 
The covariant regulator for the baryon-meson interaction is chosen to be a dipole form 
\begin{equation}
\tilde{F}(k)=\left(\frac{\Lambda^{2}-m_{\phi}^{2}}{\Lambda^{2}-k^{2}}\right)^{2},
\end{equation}
where $m_\phi$ is the meson mass. For the photon field, the regulator is the same except the meson mass is replaced by zero.
In our calculation, $\Lambda$ is the only free parameter. It is found that when $\Lambda$ is around 1.0 GeV, the obtained nucleon form factors are reasonable compared with the experimental data. The determined parameters $c_1$ and $c_2$ are listed in Table I. For both cases with or without bubble and tadpole diagrams, we can reproduce the experimental nucleon magnetic moments with proper choices of $c_i$. $c_1$ and $c_2$ will be a little smaller when bubble and tadpole diagrams are included.  

\begin{figure}[hbt]
\includegraphics[scale=0.8]{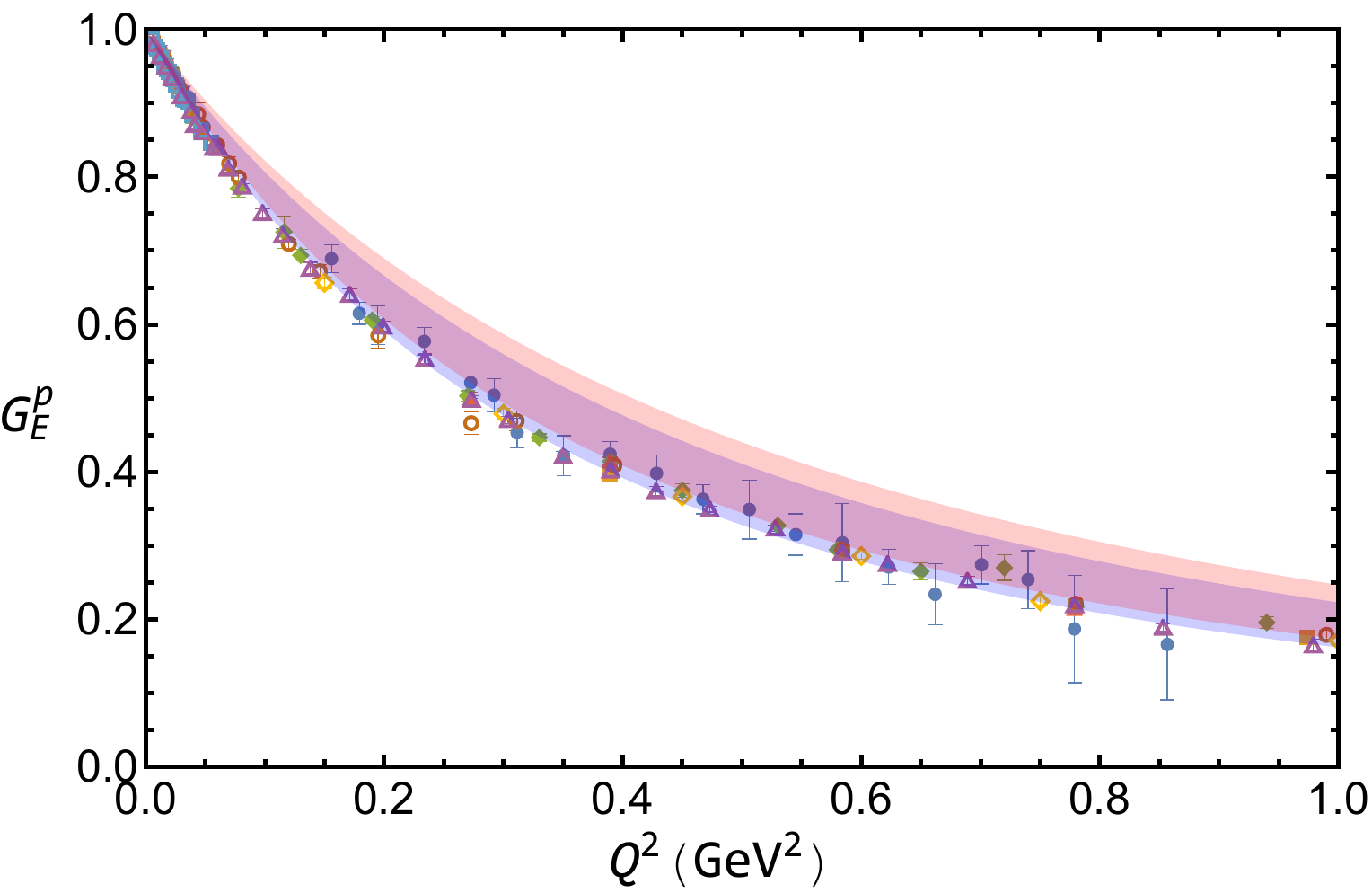}
\caption{Electric form factor of proton $G_{E}^{p}(Q^{2})$ versus momentum transfer $Q^2$ with $\Lambda=1.0 \pm 0.1$ GeV. The blue and red bands are for the results with and without bubble and tadpole diagrams, respectively.}\label{fig:gep}
\end{figure}

\begin{figure}[hbt]
\includegraphics[scale=0.8]{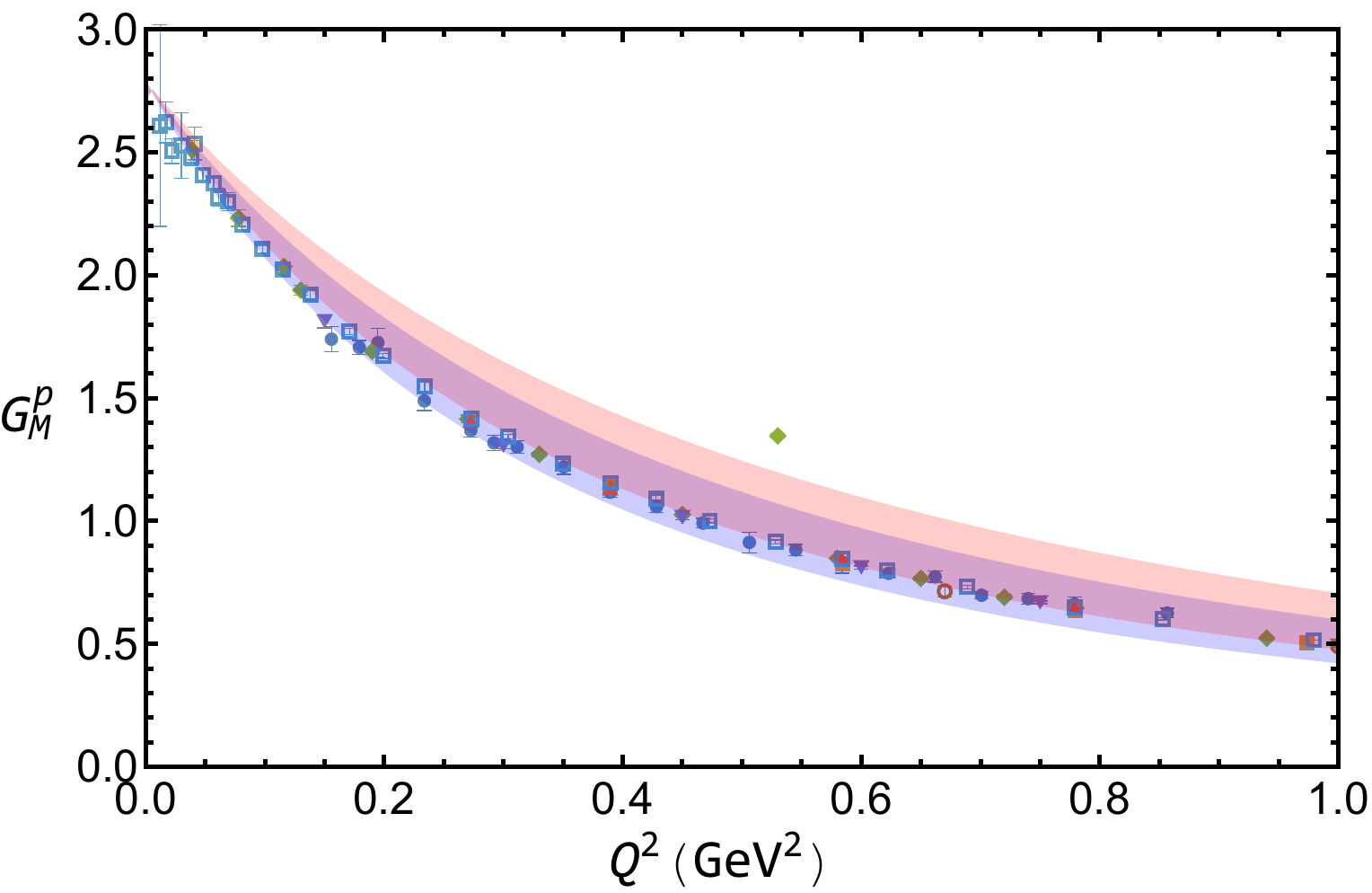}
\caption{Same as Fig.~\ref{fig:gep} but for magnetic form factor.}\label{fig:gmp}
\end{figure}

With the determined parameters, we will show the results of the electromagnetic and strange form factors of proton and neutron. The proton electric form factor versus momentum transfer is plotted in Fig.~\ref{fig:gep}. The blue and red bands are the results for $\Lambda$ $1.0 \pm 0.1$ GeV with and without bubble and tadpole diagrams, respectively. The experimental data are also shown in the figure. From the figure, one can see the experimental data can be reasonably described in both cases.
The obtained charge form factor $G_E^p(Q^{2})$ with bubble and tadpole contributions is a little smaller than that without bubble and tadpole contributions. With the additional diagram generated from the interaction of gauge link, the proton charge is 1 at $Q^2=0$. Compared with the results with dimensional and infrared regularizations, the nonlocal approach can describe the form factor data at relatively large momentum transfer.
The calculated  magnetic form factor of proton $G_{M}^{p}(Q^{2})$ is shown in Fig.~\ref{fig:gmp}. 
In both cases, the theoretical bands are also consistent with the experimental data up to 1 GeV$^2$. The band with bubble and tadpole contributions is a little smaller than that without those contributions, which is more closer to the experimental data as the electric form factor case.

\begin{figure}[hbt]
\includegraphics[scale=0.8]{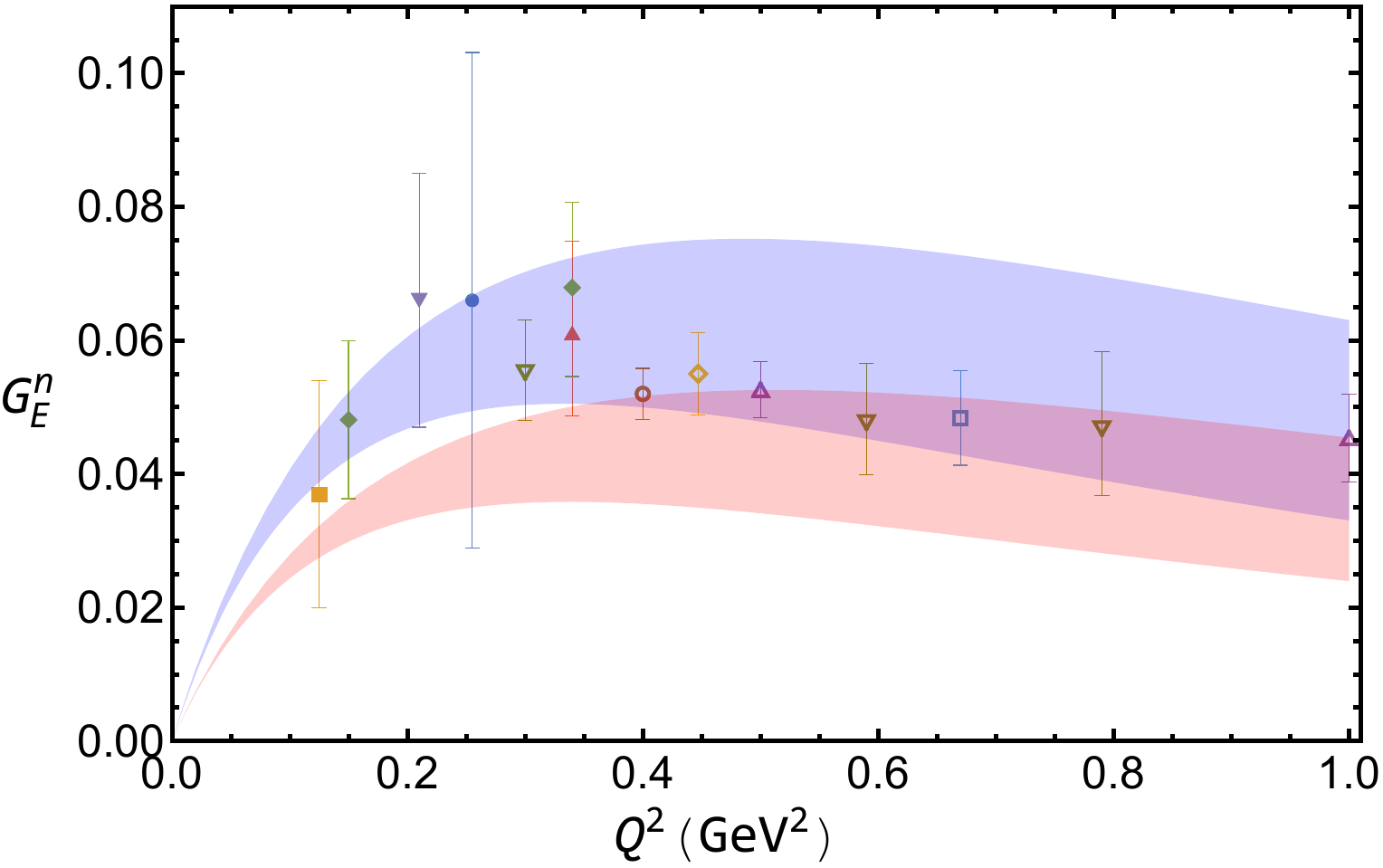}
\caption{Electric form factor of neutron $G_{E}^{n}(Q^{2})$ versus momentum transfer $Q^2$ with $\Lambda=1.0 \pm 0.1$ GeV. The blue and red bands are for the results with and without bubble and tadpole diagrams, respectively.}
\label{fig:gen}
\end{figure}

\begin{figure}[hbt]
\includegraphics[scale=0.8]{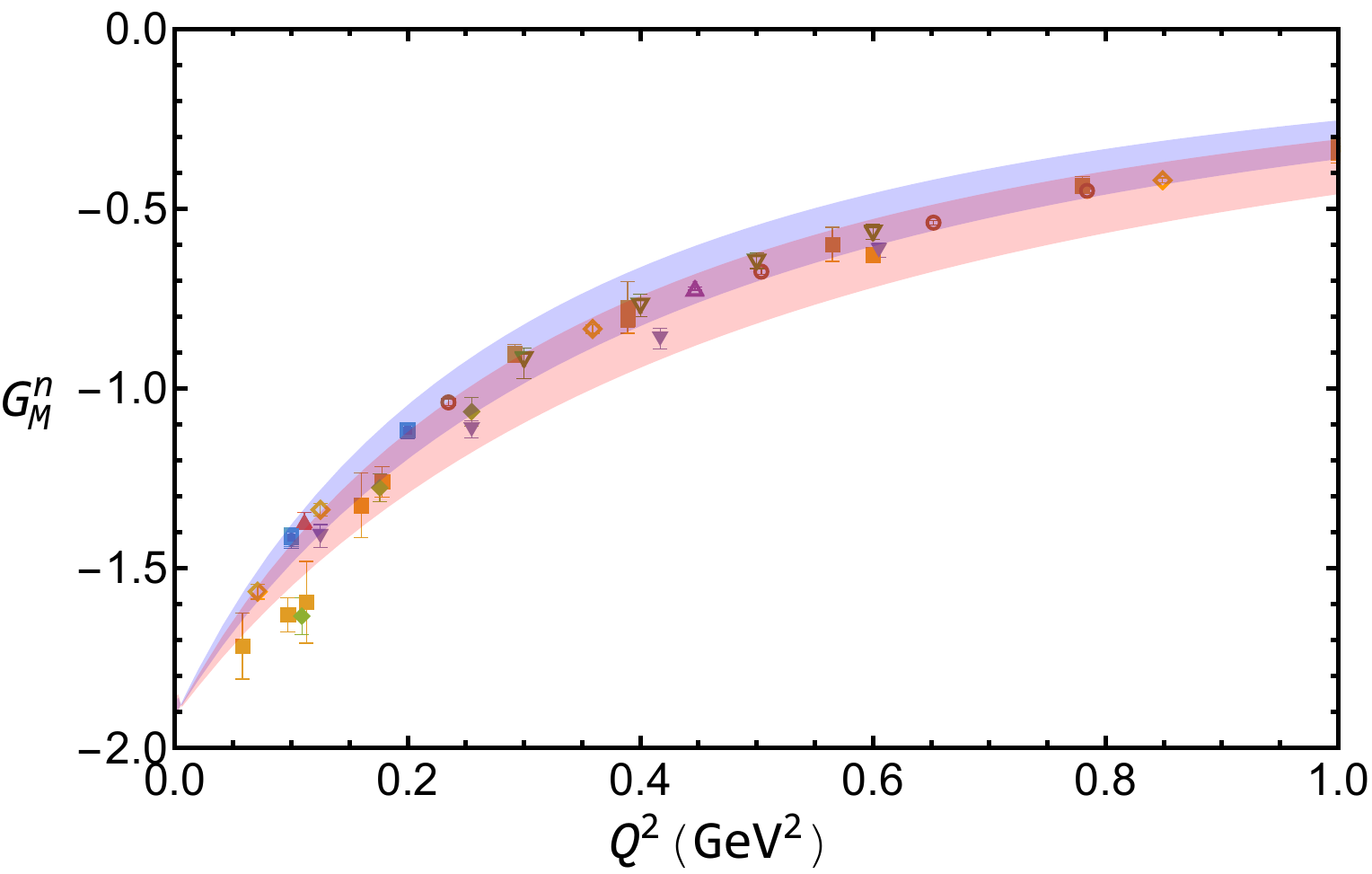}
\caption{Same as Fig.~\ref{fig:gen} but for magnetic form factor.}\label{fig:gmn}
\end{figure}

Same as the proton case, the electromagnetic form factors of neutron, $G_{E}^{n}(Q^{2})$ and $G_{M}^{n}(Q^{2})$
are plotted in Fig.~\ref{fig:gen} and Fig.~\ref{fig:gmn}. 
For $G_E^n(Q^2)$, both of the bands first increases and then smoothly decreases with the increasing momentum transfer. The shapes are both comparable with the experimental data.
The band with bubble and tadpole contributions is higher than that without those contributions. Obviously, the bubble and tadpole contributions make the theoretical result more reasonable. With the additional diagrams, the neutron charge is 0 at $Q^2=0$. We should mention that the neutron electric form factor in the case without bubble and tadpole diagrams is larger than our previous result \cite{He1}. This is mainly because the parameters are a little different. In particular, here the cutoff $\Lambda$ is chosen to be around 1.0 GeV, while it is 0.85 GeV in Ref.~\cite{He1}.  
For $G_M^n(Q^2)$, there is a little difference between these two bands, though both of them can describe experimental data well.

From the above numerical results, one can see that for the electromagnetic form factors, though the determined low energy constants $c_1$ and $c_2$ are different, the calculated results with and without bubble and tadpole contributions are both comparable with experimental data. With the determined parameters, we then present results for the strange form factors. 

\begin{figure}[hbt]
\includegraphics[scale=0.8]{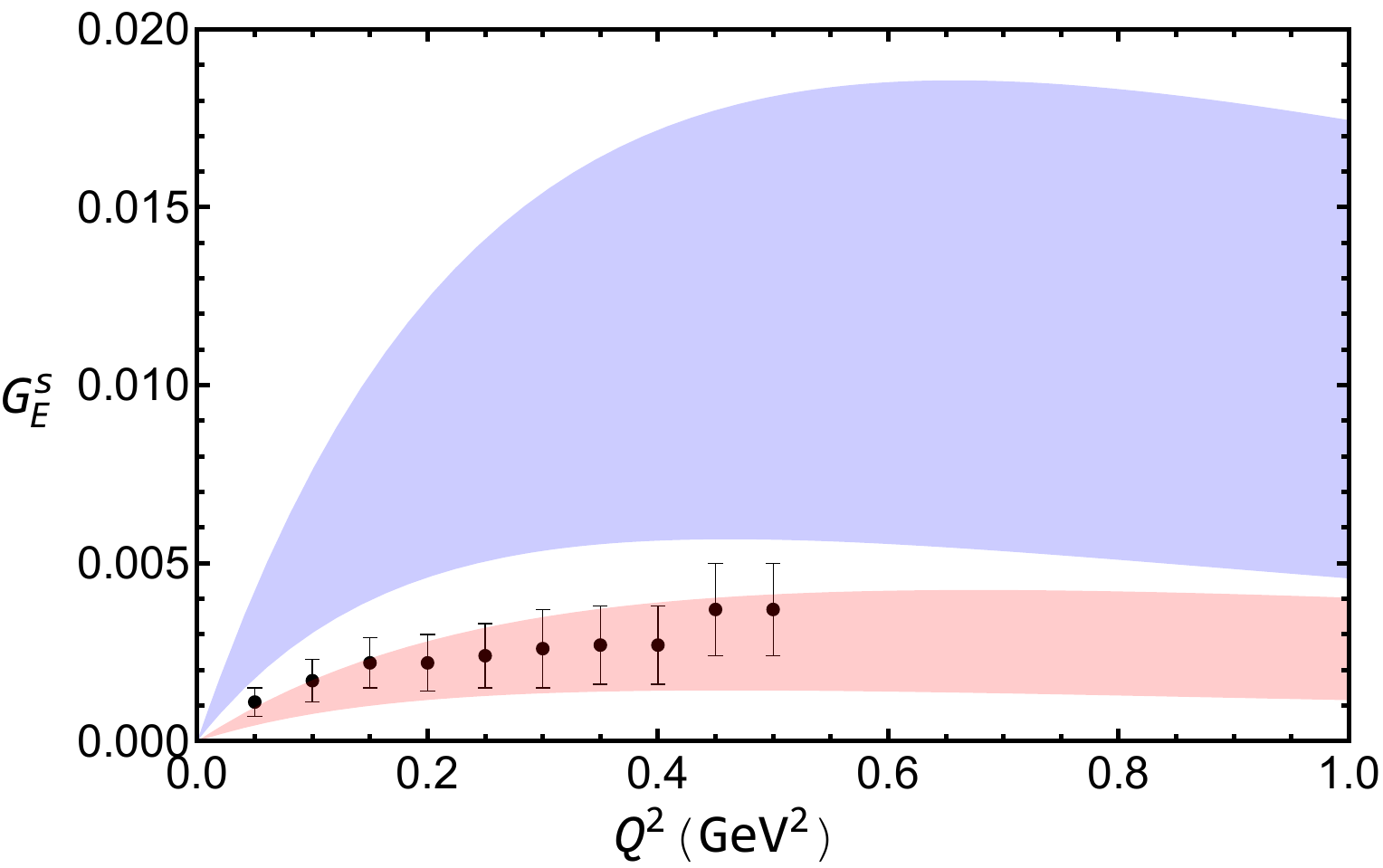}
\caption{Strange electric form factor of nucleon $G_{E}^{s}(Q^{2})$ versus momentum transfer $Q^2$ with $\Lambda=1.0 \pm 0.1$ GeV. The blue and red bands are for the results with and without bubble and tadpole diagrams, respectively. The lattice data are from Ref.~\cite{Sufian}.}\label{fig:ges}
\end{figure}

The strange electric form factor of nucleon, $G_{E}^{s}(Q^{2})$ for $\Lambda$ $1.0 \pm 0.1$ GeV is shown in Fig.~\ref{fig:ges}. 
As for the nucleon electromagnetic case, the blue and red bands are for the results with and without bubble and tadpole contributions, respectively. The lattice data from
Ref.~\cite{Sufian} are shown as well. In both cases, the strange electric form factor first increase from zero and then change smoothly with the increasing $Q^2$. The numerical difference of these two cases is obvious. The result with the bubble and tadpole contributions is much larger than that without these contributions. For example, at $Q^2=0.22$ GeV$^2$, $G_E^s$(0.22 GeV$^2$) $= 0.0085_{-0.0037}^{+0.0046}$ which is about 4 times larger than the strange electric form factor without bubble and tadpole contributions. This is because the contributions from bubble and tadpole diagrams are much larger than those from the diagrams with octet and decuplet intermediate states. We should mention that after we fixed the parameters $c_1$ and $c_1$ by the magnetic moments of proton and neutron, the strange form factors are calculated directly without any parameter. Though our values are larger than the lattice simulation at finite $Q^2$, the numbers are consistent with the global analysis where $G_E^{s,\text{exp}}(0.22~\text{GeV}^2)$ $=0.035\pm 0.030 \pm 0.019$ \cite{Baunack}.

\begin{figure}[hbt]
\includegraphics[scale=0.8]{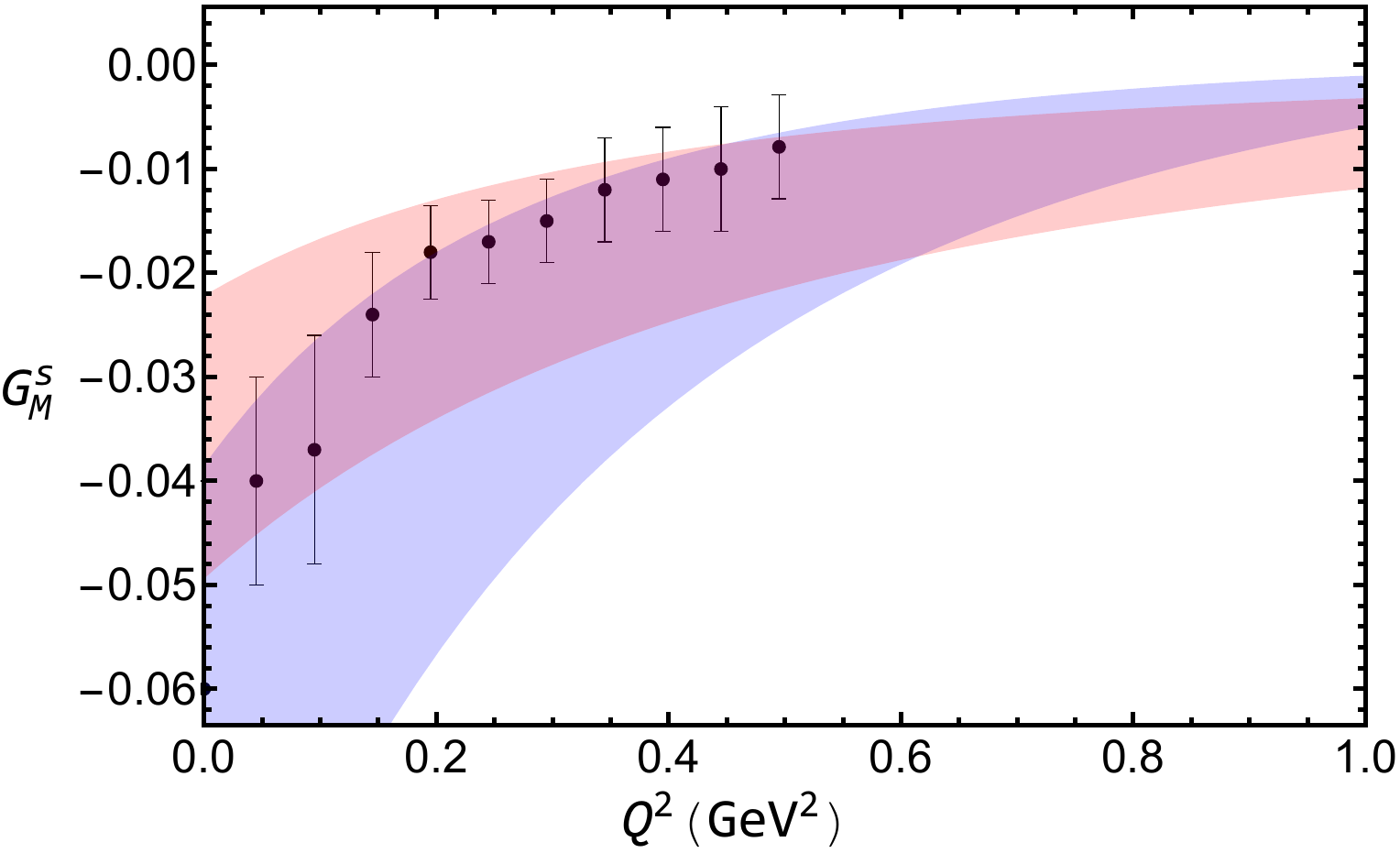}
\caption{Same as Fig.~\ref{fig:ges} but for magnetic form factor.}\label{fig:gms}
\end{figure}

The strange magnetic form factor is plotted in Fig.~\ref{fig:gms} for the two cases together with the lattice data. For the magnetic case, the contributions from bubble and tadpole diagrams are comparable with those from the rainbow and Kroll-Rudderman diagrams. As a result, the two bands are both negative and comparable. The strange magnetic moment with bubble and tadpole contribution is $\mu_s = -0.066_{-0.034}^{+0.028}$, which is also covered by the experimental analysis $\mu_s^{\text{exp}} = -0.14 \pm 0.11 \pm 0.11$ \cite{Baunack}.

\section{Summary}

We applied the nonlocal EFT to calculate the electromagnetic and strange form factors of nucleon. The bubble and tadpole diagrams are included in the calculation. The next-to-leading order baryon-meson interaction ${\cal L}^{(2)}$ is necessary which was not needed in the previous calculation with only rainbow and Kroll-Rudderman diagrams. In the numerical calculation, all the parameters are predetermined except the cutoff parameter $\Lambda$ in the regulator and the low energy constants $c_1$ and $c_2$. $c_1$ and $c_2$ are fixed by the nucleon magnetic moments and $\Lambda$ is chosen to be around 1.0 GeV. Numerical results show that the electromagnetic form factors 
with and without bubble and tadpole diagrams are close to each other with the proper choices of $c_1$ and $c_2$. 
In both cases, the electromagnetic form factors are in good agreement with the experimental data.
Since there is no free parameter when we calculate the strange form factors, the obtained strange form factors could have big difference in two cases. With the bubble and tadpole diagrams, the magnitudes of strange electric and magnetic form factors are both larger. The obtained magnetic moment is $\mu_s =-0.066_{-0.034}^{+0.028}$. The strange electric form factor is positive and $G_E^s$(0.22 GeV$^2$)$ = 0.0085_{-0.0037}^{+0.0046}$ when $Q^2=0.22$ GeV$^2$. Our results show the magnitudes of strange form factors are larger than the lattice data. 

\section*{Acknowledgments}

This work is supported by the National Natural Science Foundation of China (NSFC) under Grant No. 11975241.

\end{document}